\documentclass[useAMS]{mn2e}
\usepackage{amssymb,amsmath}
\usepackage{graphicx}
\usepackage{lscape}
\usepackage{xcolor}

\title[PG 1613+658]
{Are broad optical balmer lines from central accretion disk in PG 1613+658?}
\author[Zhang X.-G.]
       {Zhang, Xue-Guang$^{1,2}$\\
       $^1$Purple Mountain Observatory, Chinese Academy of Sciences,
             2 Beijing XiLu, NanJing, JiangSu, 210008, P. R. China \\
       $^2$Chinese Center for Antarctic Astronomy, NanJing,
             JiangSu, 210008, P. R. China}

\date{}
\def\LaTeX{L\kern-.36em\raise.3ex\hbox{a}\kern-.15em
    T\kern-.1667em\lower.7ex\hbox{E}\kern-.125emX}
\begin{document}

\pagerange{\pageref{firstpage}--\pageref{lastpage}} \pubyear{2014}
\maketitle
\label{firstpage}

\begin{abstract}
   In this letter, we report positive correlations between broad line width 
and broad line flux for the broad balmer lines of the long-term observed 
AGN PG 1613+658. Rather than the expected negative correlations under the 
widely accepted virialization assumption for AGN BLRs, the positive 
correlations indicate much different BLR structures of PG 1613+658 from 
the commonly considered BLR structures which are dominated by the equilibrium 
between radiation pressure and gas pressure. Therefore, accretion disk origin 
is preferred for the observed broad single-peaked optical balmer lines of 
PG 1613+658, because of the mainly gravity dominated disk-like BLRs with radial 
structures having few effects from radiation pressure.
\end{abstract}

\begin{keywords}
Galaxies:Active -- Galaxies:nuclei -- Galaxies:quasars:Emission lines
-- Galaxies: Individual: PG 1613+658
\end{keywords}

\section{Introduction}
    
    Broad emission line regions (BLRs) of active galactic nuclei (AGNs) so 
far can not be resolved spatially, however, many efforts have been done to 
study geometric and dynamic structures of AGN BLRs by properties of 
spectroscopic and photometric variability (Blandford \& Mckee 1982, 
Peterson et al. 1993, Bentz et al. 2010, Pancoast et al. 2011, Grier et al. 
2013, Kollatschny \& Zetzl 2011, 2013, Pancoast et al. 2013, Baskin et al. 
2014, Kollatschny et al. 2014). Moreover, under the commonly and widely 
accepted virialization assumption for AGN BLRs, properties of broad emission 
lines can be applied to conveniently estimate virial black hole masses of 
broad line AGNs (Peterson et al. 2004, Kelly \& Bechtold 2007, Krause et al. 
2011, Woo et al. 2013, Ho \& Kim 2014):
\begin{equation}
M_{BH} \propto V^2_{broad}\times R_{BLRs} 
        \propto V^2_{broad}\times \lambda L_{5100\AA}^{\sim0.5}
\end{equation}
where $V_{broad}$ represents broad line width which can be treated as the 
substitute for rotating velocities of broad line emission clouds, 
$R_{BLRs}$ means the distance between BLRs and central black hole 
(the BLRs size) which can be estimated by the empirical relation 
$R_{BLRs}\propto \lambda L_{5100\AA}^{0.5}$ (Wang \& Zhang 2003, Kaspi et al. 
2005, Bentz et al. 2013). Moreover, with the considerations of the strong 
correlation between continuum luminosity and broad line luminosity 
(Greene \& Ho 2005), the equation above can be re-written as
\begin{equation}
M_{BH} \propto V^2_{broad}\times L^{\sim0.5}_{broad}
\end{equation}
From the equation above, we can expect one strong negative correlation between 
broad line width and broad line luminosity, 
$V^2_{broad}\times L^{\sim0.5}_{broad}=constant$, for individual long-term 
observed AGN.

    Several AGNs have been reported on the expected negative correlations 
between broad line width and broad line luminosity (or BLR size) (Peterson 
et al. 2004), however, in our previous papers on 3C390.3 (Zhang 2013a) and 
PG 0052+251 (Zhang 2013b), we have reported unexpected positive correlations 
between broad line width and broad line luminosity for the double-peaked broad 
H$\alpha$ of 3C390.3 and for the intermediate broad component of optical balmer 
lines of PG 0052+521. Here, we report one another positive correlation between 
broad line width and broad line flux of broad balmer lines in PG 1613+658, 
and then give further discussions on the probable origin of broad balmer 
lines from central accretion disk. The letter is organized as follows. 
Section 2 gives our main results. Section 3 shows our discussions and 
conclusions. 

\section{Main Results}

\begin{figure}
\centering\includegraphics[width=8cm,height=7cm]{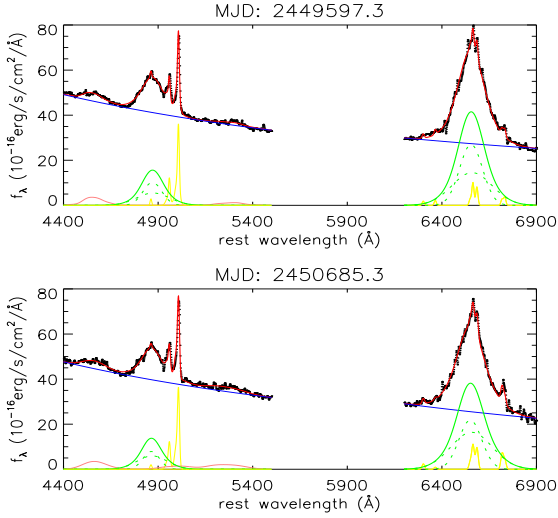}
\caption{Two examples on the best fitted results for emission lines in the 
spectra observed on 1th, Sep. 1994 and observed on 24th, Aug. 1997. In each 
panel, solid line in black represents the observed spectrum, solid line in 
red are for the best fitted results, solid line in blue is for the power law 
AGN continuum, solid lines in green, yellow and pink near the bottom are for 
the broad line, for the narrow emission lines and for the optical Fe~{\sc ii} 
respectively,  dotted lines in green are the two determined broad 
gaussian components.}
\label{fit_line}
\end{figure}

   PG 1613+658 (=PG1613) is one well known long-term observed AGN in the 
sample of Kaspi et al. (2000). There are 44 public optical spectra with both 
broad H$\beta$ and broad H$\alpha$, which can be collected from the website 
http://wise-obs.tau.ac.il/\~{}shai/PG/. The detailed descriptions about the 
observing techniques and  reduction procedures for the public spectra can 
be found in Kaspi et al. (2000). The collected spectra observed from 21th, 
May, 1991 to 16th, Sep., 1998 have been binned into 1$\AA$ per pixel and been 
padded from 3000$\AA$ to 9000$\AA$. Then, line parameters of emissions of 
PG1613 are measured as follows

   All the emission lines are fitted simultaneously within the rest 
wavelength ranges from 4400\AA to 5500\AA and from 6200\AA to 6900\AA. 
Here, for each broad balmer emission line, two broad gaussian 
functions rather than one broad gaussian function are applied, because of 
complicate broad line profile.  For optical Fe~{\sc ii} lines, the more 
recent optical Fe~{\sc ii} template in Kovacevic et al. (2010) has been 
applied in our procedure. For narrow emission lines (narrow balmer lines, 
[O~{\sc iii}]$\lambda4959,5007\AA$, [O~{\sc i}]$\lambda6300,6363\AA$, 
[N~{\sc ii}]$\lambda6548,6583\AA$ and [S~{\sc ii}]$\lambda6716,6731\AA$ 
doublets), they are described by narrow gaussian functions with similar 
line profiles, i.e., they have the same emission line redshift, the same 
line width. Furthermore, one broad gaussian function is applied for the 
much weak He~{\sc ii} line, and two extended broad components are applied for 
the extended wings of [O~{\sc iii}] doublet. And moreover, the [O~{\sc iii}] 
and [N~{\sc ii}] doublets have the fixed theoretical flux ratios expected 
by atomic physics: the flux ratios of [O~{\sc iii}]$\lambda5007\AA$ to 
[O~{\sc iii}]$\lambda4959\AA$ and [N~{\sc ii}]$\lambda6583\AA$ to 
[N~{\sc ii}]$\lambda6548\AA$ are fixed to 3 (Dimitrijevic et al. 
2007). Then, two independent power law functions are applied for the AGN 
continuum under the H$\beta$ and under the H$\alpha$. Finally, line parameters 
can be well determined through Levenberg-Marquardt least-squares minimization 
method. Figure~\ref{fit_line} shows two examples on the best 
fitted results for the emission lines, including each determined broad line  
described by two broad gaussian components.

\begin{table}
\centering
\begin{minipage}{80mm}
\scriptsize
\caption{Line Parameters}
\begin{tabular}{cccccc}
\hline
MJD & $\sigma(H\beta)$  & flux(H$\beta$) & $\sigma(H\alpha)$  & 
flux(H$\alpha$) & flux\\
\hline
48425  &  3359$\pm$109  &  21.0$\pm$0.6  &   3425$\pm$108  &   78.0$\pm$2.2  &   97.5$\pm$2.8  \\ 
48426  &  3639$\pm$123  &  20.2$\pm$0.5  &   3699$\pm$124  &   74.9$\pm$1.9  &   92.8$\pm$2.4  \\ 
48835  &  4024$\pm$143  &  21.7$\pm$0.6  &   3269$\pm$99  &   71.4$\pm$2.2  &   82.9$\pm$2.6  \\ 
49095  &  2932$\pm$89  &  17.4$\pm$0.3  &   3308$\pm$101  &   71.6$\pm$1.4  &   88.9$\pm$1.7  \\ 
49157  &  3969$\pm$140  &  19.4$\pm$0.4  &   3477$\pm$111  &   69.3$\pm$1.6  &   86.5$\pm$2.0  \\ 
49220  &  3957$\pm$139  &  18.3$\pm$0.4  &   3412$\pm$107  &   71.3$\pm$1.9  &   84.2$\pm$2.2  \\ 
49249  &  4280$\pm$157  &  23.3$\pm$0.8  &   3561$\pm$116  &   69.5$\pm$2.3  &   92.8$\pm$3.2  \\ 
49282  &  3712$\pm$126  &  23.7$\pm$0.4  &   3348$\pm$104  &   80.3$\pm$1.4  &   106.$\pm$1.9  \\ 
49426  &  3523$\pm$117  &  19.1$\pm$0.5  &   3654$\pm$121  &   77.0$\pm$2.1  &   98.5$\pm$2.7  \\ 
49453  &  3902$\pm$136  &  21.3$\pm$0.5  &   3891$\pm$136  &   74.7$\pm$1.8  &   94.6$\pm$2.3  \\ 
49487  &  3744$\pm$128  &  21.9$\pm$0.3  &   3565$\pm$116  &   73.9$\pm$1.2  &   97.5$\pm$1.6  \\ 
49519  &  4752$\pm$185  &  23.6$\pm$0.5  &   4367$\pm$168  &   85.0$\pm$1.9  &   99.5$\pm$2.3  \\ 
49541  &  4464$\pm$168  &  23.4$\pm$0.2  &   4273$\pm$161  &   82.5$\pm$1.0  &   99.0$\pm$1.2  \\ 
49569  &  4256$\pm$156  &  22.5$\pm$0.3  &   4261$\pm$160  &   83.5$\pm$1.1  &   99.0$\pm$1.4  \\ 
49597  &  4330$\pm$160  &  24.5$\pm$0.3  &   3878$\pm$135  &   79.3$\pm$1.2  &   99.1$\pm$1.5  \\ 
49597  &  4368$\pm$162  &  23.1$\pm$0.3  &   4100$\pm$149  &   85.2$\pm$1.1  &   101.$\pm$1.4  \\ 
49784  &  4792$\pm$188  &  24.6$\pm$0.5  &   4079$\pm$148  &   81.6$\pm$1.7  &   98.3$\pm$2.1  \\ 
49813  &  4806$\pm$189  &  24.5$\pm$0.3  &   4099$\pm$149  &   80.7$\pm$1.1  &   99.2$\pm$1.4  \\ 
49828  &  4413$\pm$165  &  24.1$\pm$0.2  &   4079$\pm$148  &   85.2$\pm$1.0  &   102.$\pm$1.2  \\ 
49838  &  4592$\pm$176  &  23.7$\pm$0.3  &   4192$\pm$156  &   85.0$\pm$1.1  &   101.$\pm$1.3  \\ 
49876  &  4276$\pm$157  &  23.5$\pm$0.3  &   4119$\pm$151  &   80.3$\pm$1.0  &   99.8$\pm$1.3  \\ 
49897  &  4413$\pm$165  &  22.4$\pm$0.4  &   4329$\pm$165  &   81.3$\pm$1.7  &   97.3$\pm$2.1  \\ 
49915  &  4040$\pm$144  &  22.1$\pm$0.2  &   3728$\pm$126  &   77.4$\pm$0.8  &   94.9$\pm$1.0  \\ 
49919  &  4571$\pm$174  &  21.4$\pm$0.5  &   4167$\pm$154  &   78.5$\pm$1.8  &   95.6$\pm$2.2  \\ 
49962  &  4200$\pm$153  &  20.6$\pm$0.3  &   4050$\pm$146  &   76.7$\pm$1.3  &   94.1$\pm$1.6  \\ 
49984  &  4455$\pm$168  &  23.3$\pm$0.3  &   3947$\pm$139  &   76.7$\pm$1.1  &   94.8$\pm$1.4  \\ 
49989  &  4205$\pm$153  &  22.7$\pm$0.1  &   3791$\pm$130  &   80.1$\pm$0.6  &   98.7$\pm$0.7  \\ 
50191  &  4531$\pm$172  &  23.9$\pm$0.6  &   4151$\pm$153  &   84.5$\pm$2.3  &   102.$\pm$2.8  \\ 
50198  &  4393$\pm$164  &  24.6$\pm$0.2  &   3964$\pm$141  &   86.8$\pm$0.8  &   103.$\pm$1.0  \\ 
50241  &  3989$\pm$141  &  22.1$\pm$0.4  &   3925$\pm$138  &   78.2$\pm$1.5  &   99.2$\pm$1.9  \\ 
50244  &  4089$\pm$147  &  22.9$\pm$0.3  &   3875$\pm$135  &   84.9$\pm$1.2  &   104.$\pm$1.4  \\ 
50287  &  4129$\pm$149  &  22.9$\pm$0.2  &   3894$\pm$136  &   80.2$\pm$0.9  &   98.1$\pm$1.1  \\ 
50310  &  3781$\pm$130  &  19.3$\pm$0.7  &   3890$\pm$136  &   74.6$\pm$3.0  &   93.2$\pm$3.7  \\ 
50313  &  4771$\pm$187  &  25.0$\pm$0.4  &   3809$\pm$131  &   74.5$\pm$1.3  &   93.7$\pm$1.6  \\ 
50332  &  4504$\pm$170  &  20.9$\pm$0.2  &   3979$\pm$142  &   77.4$\pm$0.8  &   94.7$\pm$1.0  \\ 
50580  &  3900$\pm$136  &  19.8$\pm$0.9  &   3583$\pm$117  &   68.0$\pm$3.3  &   81.8$\pm$4.0  \\ 
50643  &  4877$\pm$193  &  22.2$\pm$0.4  &   3957$\pm$140  &   71.6$\pm$1.4  &   88.3$\pm$1.7  \\ 
50685  &  4115$\pm$148  &  20.3$\pm$0.4  &   4177$\pm$155  &   77.7$\pm$1.7  &   95.3$\pm$2.1  \\ 
50918  &  3876$\pm$135  &  22.9$\pm$1.1  &   4546$\pm$181  &   93.3$\pm$4.7  &   103.$\pm$5.2  \\ 
50967  &  3979$\pm$141  &  23.5$\pm$0.3  &   4338$\pm$166  &   86.8$\pm$1.1  &   103.$\pm$1.3  \\ 
51001  &  4634$\pm$178  &  24.8$\pm$0.9  &   4303$\pm$163  &   84.1$\pm$3.2  &   100.$\pm$3.8  \\ 
51027  &  4489$\pm$170  &  25.6$\pm$0.6  &   4182$\pm$155  &   82.7$\pm$2.1  &   100.$\pm$2.5  \\ 
51051  &  4418$\pm$165  &  25.0$\pm$0.3  &   4213$\pm$157  &   84.5$\pm$1.1  &   101.$\pm$1.3  \\ 
51073  &  4566$\pm$174  &  24.8$\pm$0.7  &   4497$\pm$177  &   87.1$\pm$2.7  &   104.$\pm$3.2  \\ 
\hline
\end{tabular}\\
Notice: the first column is MJD-2400000, the second and the third columns are the 
line width (the second moment) in unit of ${\rm km/s}$ and the line flux in unit 
of ${\rm 10^{-15}erg/s/cm^2}$ of broad H$\beta$, the forth and the fifth columns 
are the line width and the line flux of broad H$\alpha$, the sixth column shows 
the line flux H$\alpha$ including contributions from narrow emission lines 
collected from Kaspi et al. (2000). 
\end{minipage}
\end{table}

   Then, after subtractions of the narrow emission lines, the continuum 
emission, the Fe~{\sc ii} lines and the He~{\sc ii} line, we will have the 
clear broad line profile (similar as the sum of the determined two broad 
gaussian components), and measure line parameters through the broad line 
profile. Here, second moment rather than FWHM (full width at half maximum) is 
preferred as the line width of broad lines, because the second moment is well 
defined for arbitrary line profiles and has relatively lower uncertainty 
(Fromerth \& Melia 2000, Peterson et al. 2004). second moment ($\sigma$) and 
line flux ($flux$) are calculated by 
\begin{equation}
\begin{split}
&flux = \int P_{\lambda}d\lambda \\
&\sigma^2 = \frac{\int \lambda^2\times P_{\lambda}d\lambda}{flux} - 
         (\frac{\int \lambda\times P_{\lambda}d\lambda}{flux})^2
\end{split}
\end{equation}
where $P_{\lambda}$ represents broad line profile. Then, within the rest 
wavelength range from 4400\AA to 5600\AA for broad H$\beta$ and 6000\AA to 
7200\AA for broad H$\alpha$, the line parameters can be well determined for 
broad lines of PG1613, which are listed in Table 1.

   Moreover, corresponding uncertainties of line parameters are calculated 
as follows. Because only wavelength and flux information are included in the 
collected spectra of PG1613, it is hard to determine more accurate uncertainties 
for broad line parameters. Therefore, the reported parameter uncertainties  in 
Kaspi et al. (2000) have been applied. Based on the reported values and 
corresponding uncertainties of broad line flux (including contributions from 
narrow emission lines) in Kaspi et al. (2000) $f_{k00}\pm ferr_{k00}$, our 
measured broad line flux uncertainties are determined by 
$f_{broad} * ferr_{k00}/f_{k00}$, where $f_{broad}$ represents our measured 
broad line flux after narrow lines being subtracted. Then, similar as 
estimations of uncertainties of broad line flux, the uncertainty in FWHM given 
in Kaspi et al. (2000) is applied to estimate the uncertainty for the second 
moment of broad line.

\begin{figure}
\centering\includegraphics[width=7cm,height=3cm]{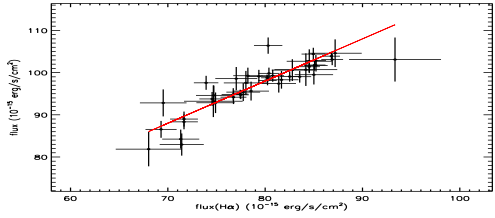}
\caption{On the correlation between line flux of broad H$\alpha$ ($flux(H\alpha)$) 
measured from line profile of broad H$\alpha$ and the reported line flux of 
H$\alpha$ ($flux$) in Kaspi et al. (2000) including contributions from narrow 
lines. The red line shows the correlation $flux = flux(H\alpha) + 18$.
}
\label{flux12}
\end{figure}

\begin{figure}
\centering\includegraphics[width=8cm,height=9cm]{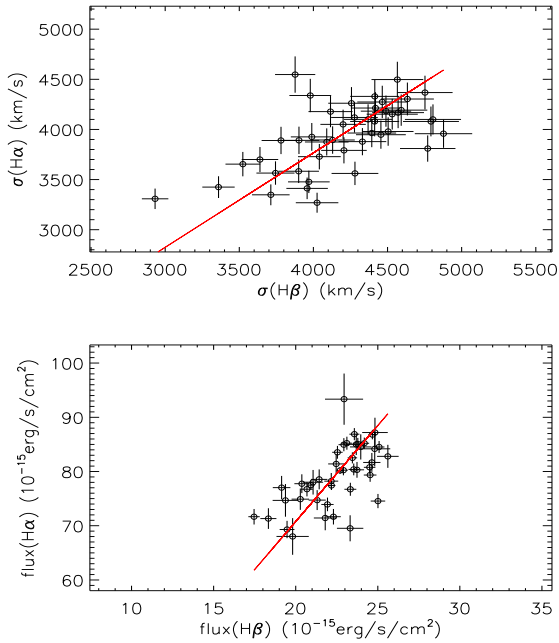}
\caption{Top panel shows the line width correlation between broad H$\alpha$ 
and broad H$\beta$. In the panel, solid line in red represents 
$\sigma(H\alpha) = 0.95\times\sigma(H\beta)$.  Bottom panel shows the line 
flux correlation between broad H$\alpha$ and broad H$\beta$. In the panel, 
solid line in red shows $flux(H\alpha) = 3.54\times flux(H\beta)$.
}
\label{Hab}
\end{figure}

\begin{figure}
\centering\includegraphics[width=8cm,height=9cm]{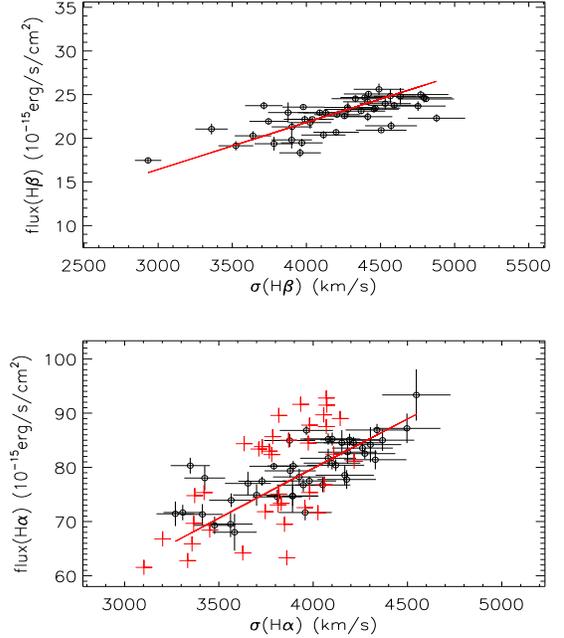}
\caption{Correlations between line width and line flux for broad H$\beta$
(top panel) and for broad H$\alpha$ (bottom panel) respectively. In the
panels, open circles are for line parameters from observed line profiles, solid
lines in red are the corresponding best fitted results. In the
bottom panel, symbol of plus in red is for the value from model expected broad
H$\alpha$ with accretion disk origin.}
\label{fs}
\end{figure}

   Then, based on measured line parameters of broad lines, we show correlations 
of broad line parameters for PG 1613. First and foremost, Figure~\ref{flux12} 
shows the correlation between the broad H$\alpha$ line flux ($flux(H\alpha)$) 
measured from line profile of broad H$\alpha$ and the reported H$\alpha$ line 
flux ($flux$) in Kaspi et al. (2000) including contributions from narrow lines. 
The spearman rank correlation coefficient is 0.92 with $P_{null}<5\times10^{-7}$. 
The strong linear correlation and the best fitted result 
$flux = flux(H\alpha) + 18$ indicate our fitted results for the broad H$\alpha$ 
have high confidence levels. Therefore, there are no further discussions on 
the spectral flux calibrations or on the effects of narrow lines on our 
following results. Besides, Figure~\ref{Hab} shows the line parameter 
correlations between broad H$\alpha$ and broad H$\beta$. Top 
panel shows the line width correlation between broad H$\alpha$ and broad 
H$\beta$.  Bottom panel shows the line flux correlation between broad H$\alpha$ 
and broad H$\beta$. The spearman rank correlation coefficients are 0.59 with 
$P_{null}\sim10^{-5}$ and 0.67 with $P_{null}\sim10^{-6}$ for the broad line 
width correlation and for the broad line flux correlation respectively. The 
strong correlations further support the high confidence levels for our 
measured broad line parameters to some extent. Last but not least, 
Figure~\ref{fs} shows the correlation between broad line width and broad line 
flux for broad balmer lines. The spearman rank correlation coefficients are 
0.67 with $P_{null}\sim10^{-6}$ and 0.76 with $P_{null}\sim10^{-9}$ for the 
correlations by parameters of broad H$\beta$ and by parameters of broad 
H$\alpha$ respectively. Then, with considerations of the uncertainties in 
both coordinates, the results can be well described as
\begin{equation}
\begin{split}
\log(flux(H\beta))&\propto (0.99\pm0.07)\times\log(\sigma(H\beta)) \\
\log(flux(H\alpha))&\propto (0.92\pm0.08)\times\log(\sigma(H\alpha)) 
\end{split}
\end{equation}.
It is clear that the results are not consistent with the expected result 
$\sigma^2\times flux^{0.5}\sim constant$ under the virialization assumption 
for AGN BLRs.

\section{discussions and conclusions}

    It is interesting to discuss what BLR structures should lead to the 
unexpected positive correlations between broad line width and broad line flux 
shown in Figure~\ref{fs}. In our previous papers on 3C390.3 (Zhang 2013a) and 
on PG0052+251 (Zhang 2013b), broad lines coming from central accretion disks 
are considered for the positive correlations. Here, in the letter, some further 
discussions on the BLR structures should be given.

   As commonly discussed BLRs, such as the more recent discussions in Baskin 
et al. (2014) (and references therein), the radiation pressure has an important 
role on radial structures of AGN BLRs. And moreover, the model in Baskin et 
al. (2014) makes definite predictions that are easily tested with just the 
flux and width measurements of the broad lines. We should test the model 
as follows. Through the equilibrium between radiation pressure and gas 
pressure, AGN BLR size sensitively depends on continuum luminosity (also on 
broad line luminosity, due to the strong correlation between continuum 
luminosity and broad line luminosity). Therefore, the flux 
weighted BLR size being increased with continuum being stronger leads to 
one strongly expected  negative correlation between broad line width and 
broad line luminosity. In order to find unexpected strong positive 
correlations for broad lines, the other model on the BLR structures 
should be considered, at least there are no or few effects from the 
equilibrium between the radiation pressure and the gas pressure on radial 
structures of BLRs. Therefore, broad lines with accretion disk origins 
(Eracleous et al. 1995, Storchi-Bergmann et al. 2003, Strateva et al. 2003, 
Lewis et al. 2010, Zhang 2011, 2013c) are firstly considered, 
because of the broad lines from gravity dominated BLRs with their radial 
structures having few effects of the radiation pressure. For the BLR 
expected by the accretion disk model, the black hole gravity also 
dominate the dynamics of emission line clouds in BLRs, however, no 
equilibrium can be commonly expected between radiation pressure and 
gas pressure for the BLRs in accretion disk, because of much smaller 
ratio of radiation pressure to gas pressure than 1 expected by standard 
Shakura-Sunyaev accretion disk. Then, we check whether such as BLR into 
accretion disk can lead to positive correlation between broad line width 
and broad line flux.

     Before proceeding further, some basic parameters of PG1613 should 
be given. For PG1613, the black hole mass is about 
$2.8\times10^8{\rm M_{\odot}}$ reported in Peterson et al. (2004). And 
moreover, the BLR size is about 40 light-days  determined by the 
reverberation mapping technique (Kaspi et al. 2000, Peterson et al. 2004). 
Then, We try to fit the broad balmer lines by the accretion disk origins, 
however, due to the single-peaked line profile without apparent double 
peaks nor clear shoulders, it is hard to find the unique solution. So that, 
the following simple disk parameters (more detailed descriptions on disk 
parameters can be found in Eracleous et al. 1995) are accepted to 
describe the observed broad H$\alpha$: inner boundary $r_0=700 R_G$, 
outer boundary $r_1=8000 R_G$ (the extended size of the disk-like BLR 
about 116 light-days), eccentricity $e=0.3$, inclination angle 
$i=67\degr$, orbital phase angle $\phi_0=-4\degr$, emissivity power 
slope $q=2$ ($f(r)\propto r^{-q}$) and local broadening velocity 
$\sigma_{loc}=1200{\rm km/s}$. Because the disk parameters lead to one 
single peaked line profile, much similar as the observed broad H$\alpha$, 
we do not show the line profile expected by the elliptical accretion 
disk model (Eracleous et al. 1995) any more. However, It is clear that 
the disk parameters lead to one clear elliptical disk-like BLR into the 
central accretion disk of PG1613, through the detailed BLR structures, 
we can try to check the correlation between broad line width and broad 
line flux as follows.      

   And moreover, some simple discussions on accretion disk size are 
shown as follows. Hawkins (2007, 2010) have shown that optical continuum 
emission regions in AGN accretion disks should be less than 10 light-days, 
which is smaller than the extended size (116 light-days) of BLR and much 
smaller than the BLR size about 40 light-days of PG1613. However, we have 
known that for double-peaked emitters, an ion torus or a hot corona around 
inner accretion disk is definitely needed for illumination on the regions 
which produce double-peaked emission lines (the energy budget problem for 
double-peaked emitters, Eracleous et al. 2003). Therefore, one  extra 
illumination source is necessary for PG1613, if the accretion disk origin 
is accepted for the broad lines, otherwise the longer distance from central 
black hole should lead to apparent energy budget problem.

   Due to the much longer relativistic precession period of the accretion 
disk (about 230 years for inner regions of the disk-like BLR) proposed in 
PG1613, effects of disk precession on line profile variability can be totally 
ignored.  Then, it is convenient to study variability of broad line having 
accretion disk origin, with considerations of continuum variability. 
Here, we make the simplifying assumption that the continuum 
emission propagates freely and isotropically in the central region. And 
moreover, we accepted that  once one hydrogen cloud captures ionization 
photons, broad line emissions are in coinstantaneous {\bf linear} response 
to ionizing continuum emissions, due to much smaller recombination time 
scale and much smaller resonance photon diffusion time scale (Peterson 
et al. 1993), and due to the strong linear correlation between continuum 
luminosity and broad line luminosity (Greene \& Ho 2005, Zhang 2014). 
Then, in order to show more clear descriptions and discussions on the 
following results, the BLR with extended size about 116 light-days is well 
evenly divided into 1600 tiny regions: the radius is 
evenly divided into 40 bins, $r_{0}\le r_{\star,i}(i=0, \dots, 40)\le r_{1}$, 
and the orbital phase angle $\phi$ is evenly separated into 40 bins,
$0\le \phi_{\star,j}(j=0\dots 40)\le 2\times\pi$. Then, the observed broad 
line is composed of the emissions from the 1600 tiny areas. The line emission 
from each tiny area can be calculated by the elliptical accretion disk 
model ($H(model)$), 
\begin{equation}
flux_{i=0,\dots,39,j=0,\dots,39} = 
   \int\limits_{r_{\star,i}}\limits^{r_{\star,i+1}}\int
      \limits_{\phi_j}\limits^{\phi_{j+1}}H(model)drd\phi
\end{equation}
Then, going with the continuum emission propagating through the extended BLR, 
the broad line profiles should be varying,
\begin{equation}
\begin{split}
flux(t) & = \sum\limits_{i=0}^{39}\sum\limits_{j=0}^{39}flux_{i,j}(t) \\
flux_{i,j}(t) & = flux_{i, j}(t-t0)\times (\frac{C_{i,j}(t)}{C_{i,j}(t-t0)})
\end{split}
\end{equation}
where $C_{i,j}(t)$ and $flux_{i,j}(t)$ represents continuum and flux intensity 
for the tiny region with $r = r_{\star,i}$ and  $\phi = \phi_j$ at time $t$, 
and $t0$ means one free time lag.  

    Then, based on continuum variability of PG1613 reported in Kaspi et al 
(2000) and the detailed BLR structure defined by the elliptical accretion 
disk parameters above, it is convenient to check the broad line variability 
by the equation (6), based on the known $flux_{i,j}(t)$ at $t=0$. Here, we 
have been accepted that at starting time $t=0$, the same continuum intensity 
is for $C_{i,j}(t)$, which has few effects on model expected results but leads 
to more convenient procedure. Moreover, one time step about 4 days ($t0=4$) 
is applied in our procedure, and the procedure is stopped, once $t$ larger than 
116 (the extended size of the BLR).  Then, the correlation between width and 
flux of model expected  broad lines (about 40 data values) is shown in 
Figure~\ref{fs}. It is clear that there is one strong positive correlation 
between broad line width and broad line flux through the simple procedure 
above. The coefficient is about 0.68 with $P_{null}\sim10^{-6}$. Furthermore, 
we can find, if model expected line parameters are used, there should be 
$\log(flux)\propto 1.22\times\log(\sigma)$, the some large slope value 1.22 
than the value listed in Equation (4) maybe due to the not well confirmed disk 
parameters. It is clear that the accretion disk origin can be applied to 
well naturally explain the unexpected positive correlation between broad 
line width and broad line flux for PG1613.

   Besides the accretion disk origins for broad lines which indicate 
one totally different BLR structures from the commonly considered BLR 
structures, there are some other cases for the BLR with less importance 
of the equilibrium between the radiation pressure and the gas pressure, 
such as the radial flow structures in the common BLRs. However, due 
to the few contributions of the radial flows to the BLR size and to 
the total broad line width which are dominated by the gas pressure 
and radiation pressure, it is hardly to find the positive correlation 
shown in the Figure~\ref{fs} due to the radial structures. Therefore, 
in the letter, there are no further discussions on the subtle 
structures in common AGN BLRs.

   Before the end of the letter, we compare the positive correlations 
among the double-peaked emitter 3C390.3, the normal QSO PG 0052+251 
and PG 1613+658. Under the mathematical formula 
$\log(flux)\propto\alpha\times\log(\sigma)$, the slope values are 
$\alpha\sim0.47\pm0.08$, $\alpha\sim1.83\pm0.12$ and 
$\alpha\sim0.96\pm0.08$ for the double-peaked broad H$\alpha$ of 
3C390.3, for the intermediate broad optical balmer lines of PG0052+251 
and for the broad optical balmer lines of PG1613+658 respectively. 
The different slopes probably indicate some different 
disk-like BLR physical parameters, which will be studied in detail 
in one following being prepared manuscript.

   Although the detailed BLR structures are still unclear for AGN, 
some basic properties of BLRs can be applied to anticipate the 
observed broad line properties. Under the commonly accepted 
virialization assumption and the basic radial BLR structures for 
the vast majority of broad line AGNs, the strong negative correlation 
can be expected between broad line width and broad line flux. 
However, as one well theoretical model defined disk-like BLR for 
the broad lines with accretion disk origins, the totally gravity 
dominated BLR can do naturally lead to the strong positive 
correlation between broad line width and broad line flux. In other 
words, the strong positive correlation for the broad line can be 
used as one probable indicator for the AGN broad lines with 
accretion disk origins.

\section*{Acknowledgements}
Zhang, X.-G. very gratefully acknowledge the anonymous referee for 
giving us constructive comments and suggestions to greatly improve our 
paper.  ZXG gratefully acknowledges the support from  
NSFC-11003043 and NSFC-11178003, and gratefully thanks Dr. Kaspi S. to 
provide public observed spectra of PG1613+658. 
(http://wise-obs.tau.ac.il/\~{}shai/PG/).

\label{lastpage}
\end{document}